\documentclass[preprint,12pt]{elsarticle}

\usepackage{amssymb}
\journal{Annals of Physics}

\begin{document}

\begin{frontmatter}



\title{A New Method for Multi-Bit and Qudit Transfer\\ Based on Commensurate Waveguide Arrays}


\author[Serbia] {J. Petrovic\corref{cor1}}
\ead{jovanap@vin.bg.ac.rs}\cortext[cor1]{Corresponding author}
\author[US,Greece]{J. J. P. Veerman}
\address[Serbia]{Vinca Institute of Nuclear Sciences, University of Belgrade, 12-14 Mike Alasa, 11000 Belgrade, Serbia}
\address[US]{Fariborz Maseeh Dept. of Math. and Stat., Portland State Univ., Portland, OR, USA}
\address[Greece]{CCQCN, Dept of Physics, University of Crete, 71003 Heraklion, Greece}

\begin{abstract}
The faithful state transfer is an important requirement in the construction of classical and quantum computers. While the high-speed transfer is realized by optical-fibre interconnects, its implementation in integrated optical circuits is affected by cross-talk. The cross-talk between densely packed optical waveguides limits the transfer fidelity and distorts the signal in each channel, thus severely impeding the parallel transfer of states such as classical registers, multiple qubits and qudits. Here, we leverage on the suitably engineered cross-talk between waveguides to achieve the parallel transfer on optical chip. Waveguide coupling coefficients are designed to yield commensurate eigenvalues of the array and hence, periodic revivals of the input state. While, in general, polynomially complex, the inverse eigenvalue problem permits analytic solutions for small number of waveguides. We present exact solutions for arrays of up to nine waveguides and use them to design realistic buses for multi-(qu)bit and qudit transfer. Advantages and limitations of the proposed solution are discussed in the context of available fabrication techniques.

\end{abstract}


\end{frontmatter}


\section{Introduction}
\label{sec:Intro}
The faithful transfer of information from one specified location to another is a necessary condition for error-free computation. Simultaneously, the requirement for high transfer speed has directed hardware development towards optical technologies. The long-haul parallel transfer is traditionally realized via optical fibres, while the integration of photonic and semiconductor technologies has recently put the integrated optical waveguides at the forefront of the interconnect research \cite{InterconnectsOX15}. However, a-few-to-several-micron diameter of guided modes causes significant cross-talk between densely packed waveguides that prevents a straightforward extension to a large number of communication channels \cite{HaurylauJSTQE06}. Proposed solutions rely on high refractive-index contrast waveguides, wavelength-division multiplexing \cite{HaurylauJSTQE06}, superlattices \cite{SungNatComms15} and the use of supermodes of a multicore fibre \cite{SercanPTL13}. The reduction in cross-talk has been achieved at the cost of propagation delay due to the high refractive index, complexity of multiplexing components, fanning out of waveguides at the input and output of superlattices, and strict requirements on radius and spacing of the fibre cores, respectively.

Optical technology is likewise a promising solution for the transfer of quantum states in future quantum computers where the coherent transfer of state superpositions, hence of both amplitude and phase, is mandatory \cite{UnruhPRA95, DiVincenzoArxiv00}. Photons naturally act as flying qubits and permit information encoding from atoms \cite{MaxeinNP13} and quantum dots \cite{AndrewPRL15}. The transfer of a single qubit encoded in photon polarization can be easily performed by sending a photon through a polarization-maintaining fibre or waveguide. Mirroring of the input state amplitude at the output of an array, known as perfect transfer \cite{BosePRL03}, has also emerged as a feasible strategy and has been demonstrated in an optical waveguide array (WGA) \cite{ChapmanNatComms16}. However, even the simplest universal quantum gates operating on two qubits and an auxiliary signal require a five-state Hilbert space. Generally, the number of channels needed for the transfer of $n$ qubits scales as $2^n$, \cite{ShoreArxiv05}. If a qudit representation is used instead \cite{KiktenkoPLA15}, the number of channels scales linearly with the number of states. The scalability requirement is challenging for the methods suggested for the single-qubit transfer and is further convoluted by the cross-talk in densely packed arrays.

Here, we propose a new method and optical hardware for the parallel transfer of classical bits, qubits and a qudit that leverages on cross-talk between waveguides. The method is based on the full state revivals in linearly coupled commensurate waveguide arrays. Periodic dynamics are achieved by imposing a condition of commensurability of all eigenvalues of the array. Whereas arrays with $n<4$ waveguides are always commensurate, the eigenvalues of longer finite arrays are commensurate only for certain ratios of their coupling coefficients. The key engineering challenge is to find these ratios, which are then easily converted into the ratios of interwaveguide separations as a design parameter \cite{BellecOL12}. It belongs to a class of inverse eigenvalue problems that have analytical solutions in a small number of cases and are, in general, of polynomial complexity \cite{inverse}. Analytical solutions have been reported for short mirror-symmetric commensurate waveguide arrays employed as directional couplers \cite{PetrovicOL15}. Here, we present the solutions for arrays with up to 9 waveguides and use them to design parallel buses for optical computers.

In Section \ref{sec:Method}, we describe the model and present exact solutions of the inverse eigenvalue problem. Obtained coupling coefficients are then used to design realistic WGAs. In Section \ref{sec:Results}, we give examples of the full state revivals  transfer of multiple bits, qubits and a qudit. WGAs' capability of the perfect transfer of a single qubit is also considered. All results are corroborated by numerical simulations of realistic waveguides. In section \ref{sec:Discussion}, we discuss the impact of fabrication imperfections and beyond-the-next-neighbour coupling on the transfer fidelity, scaling of the number of qubits with the number of waveguides and preparation of input states. In the final Section \ref{sec:Conclusions}, we give conclusions and outline other configurations of WGAs of interest.

\section{Method}
\label{sec:Method}

\begin{figure}[ht!]
\begin{center}
\begin{tabular}{cc}
\includegraphics[width=5cm]{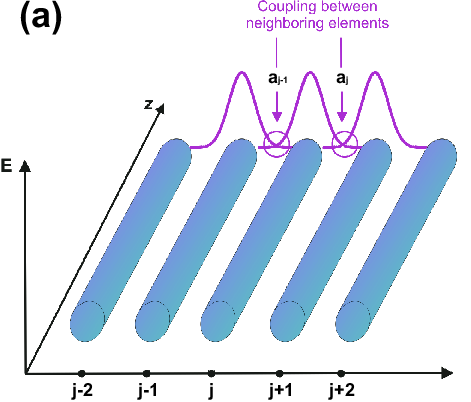}&
\includegraphics[width=5cm]{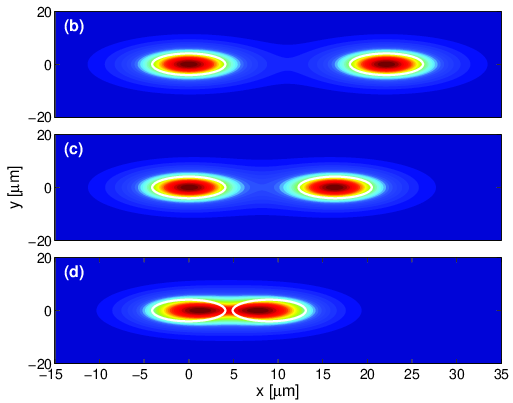}
\end{tabular}
\caption{(a) A waveguide array. The electric field amplitudes $|E|$ of supported modes (purple lines) show evanescent field overlap responsible for coupling. Coefficients $a_i$ build the coupling matrix $\mathbf{A}$ from eq.~(\ref{eq:Model}). (b) - (d) The intensity ($I\propto |E|^2$) profile of the fundamental mode of a pair of circular waveguides at the centre-to-centre distance 5.4 R (b), 4.0 R (c) and 2.2 R (d), where R is the waveguide radius. Red colour corresponds to the maximum and blue colour to the minimum intensity. White contours show waveguide cross sections.}
\label{fig:WGA}
\end{center}
\end{figure}

In a discrete approximation, waveguide cross section is assumed to be infinitesimally small allowing for the mathematical description of a mode by a complex wavefunction invariant in transversal direction. Light propagation through a discrete WGA composed of $n$ waveguides coupled linearly to their nearest neighbours is modelled by the Schr\"{o}dinger equation in the form
\begin{equation}
i\frac{\partial\psi(z)}{\partial z}= A_n\psi(z)
\label{eq:Model}
\end{equation}
where $\psi(z)=(\psi_1(z), \psi_2(z),\ldots,\psi_j(z),\ldots,\psi_n(z))$ is a complex vector state composed of wavefunctions $\psi_j(z)=|\psi_j(z)|e^{i\phi_j(z)}$ describing the light propagation through $j^{th}$ waveguide. $A_n$ is a real $nxn$ tridiagonal symmetric matrix that corresponds to the nearest-neighbour coupling \cite{Yariv}. We assume that there are no loss nor gain along the array, hence that the power of a vector state can be normalized as $\sum|\psi_j(z)|^2=1$.

\subsection{Analytic solution of the inverse eigenvalue problem}\label{subsec:Analytic}
Periodic light propagation through an n-waveguide array allows for the full state revivals and hence the transfer of n complex numbers to a distance equal to an integer number of periods. Periodic dynamics occur under the condition that all eigenvalues of the coupling matrix $\omega_j,\, j=\overline{1,n}$ are commensurate \cite{BosePRL03, BoseCP07}, which is fulfilled if there are integers $n_j$ such that
\begin{equation}
\forall\;j\;:\; \frac{L}{2\pi}\omega_j=n_j.
\label{eq:Condition}
\end{equation}
The period of oscillation is defined as the smallest $L$ for which the above equation holds for all eigenvalues of the array. In the derivations below, we express the period in terms of phase. In all cases the period is $\frac{2\pi}{q}$.

Two and three element arrays have unconditional periodicity. Their solutions are well-known and heavily applied in atomic and laser physics, the most prominent examples being Rabi coupling of nuclear spins by an oscillating magnetic field \cite{RabiPR38} and coupling of atomic levels by a resonant laser \cite{ShorePRA77}. Arrays with more than three elements generally have incommensurate eigenvalues, which leads to the propagation dynamics quasi-periodic to a certain degree \cite{CuberoPRL14}.

In what follows, we propose an inverse eigenvalue method for finding the sets of coupling coefficients that guarantee periodic propagation. We first explain a design procedure on the 4-element array, the shortest array that permits incommensurate eigenvalues, and then consider various n-dimensional cases of eq.~(\ref{eq:Model}). Since the inverse solutions for coupling coefficients are not unique, we provide general solutions with free parameters that can be specified to target a particular transfer dynamics \cite{KayIJQI10}. Detailed formal derivation of coupling coefficients can be found in Supplementary information.

\subsubsection{4-guide array}\label{subsec:A4}
We start from a coupling matrix with real nearest-neighbour interaction acting on $\mathbb{R}^4$ of the general form:
\begin{equation}\label{eq:A4}
A_4(a_1,a_2,a_3)\equiv
\left[ \begin {array}{cccc} 0&{\it a_1}&0&0\\ \noalign{\medskip}{\it
a_1}&0&{\it a_2}&0\\ \noalign{\medskip}0&{\it a_2}&0&{\it a_3}
\\ \noalign{\medskip}0&0&{\it a_3}&0\end {array} \right].
\end{equation}
Its eigenvalues $k_{1,2,3,4}$ of $A_4$ are given by:
\begin{equation}
\left[ \begin {array}{c} \frac 12 \sqrt {2{{\it a_3}}^{2}+2{{\it a_2}}^
{2}+2{{\it a_1}}^{2}+2\sqrt {{{\it a_3}}^{4}+2{{\it a_3}}^{2}{{\it
a_2}}^{2}-2{{\it a_3}}^{2}{{\it a_1}}^{2}+{{\it a_2}}^{4}+2{{\it a_2}}^
{2}{{\it a_1}}^{2}+{{\it a_1}}^{4}}}\\ \noalign{\medskip}-\frac 12 \sqrt {2
{{\it a_3}}^{2}+2{{\it a_2}}^{2}+2{{\it a_1}}^{2}+2\sqrt {{{\it
a_3}}^{4}+2{{\it a_3}}^{2}{{\it a_2}}^{2}-2{{\it a_3}}^{2}{{\it a_1}}^{
2}+{{\it a_2}}^{4}+2{{\it a_2}}^{2}{{\it a_1}}^{2}+{{\it a_1}}^{4}}}
\\ \noalign{\medskip}\frac 12 \sqrt {2{{\it a_3}}^{2}+2{{\it a_2}}^{2}+2
{{\it a_1}}^{2}-2\sqrt {{{\it a_3}}^{4}+2{{\it a_3}}^{2}{{\it a_2}}^
{2}-2{{\it a_3}}^{2}{{\it a_1}}^{2}+{{\it a_2}}^{4}+2{{\it a_2}}^{2}{{
\it a_1}}^{2}+{{\it a_1}}^{4}}}\\ \noalign{\medskip}-\frac 12 \sqrt {2{{
\it a_3}}^{2}+2{{\it a_2}}^{2}+2{{\it a_1}}^{2}-2\sqrt {{{\it a_3}}^
{4}+2{{\it a_3}}^{2}{{\it a_2}}^{2}-2{{\it a_3}}^{2}{{\it a_1}}^{2}+{{
\it a_2}}^{4}+2{{\it a_2}}^{2}{{\it a_1}}^{2}+{{\it a_1}}^{4}}}
\end {array} \right].
\end{equation}

\noindent We can simplify the above expressions by writing
\begin{equation}
\begin {array}{ccc}\label{eq:xu}
x^2 &\equiv& a_1^2+a_2^2+a_3^2\\
u^2 &\equiv & a_1a_3
\end{array}
\end{equation}
which gives for the eigenvalues
\begin{equation}
\left[\begin{array}{ccc}
\pm \frac 12 \sqrt{2x^2 \pm 2\sqrt{x^4-4u^4} }
\end{array}\right].
\end{equation}
From (\ref{eq:Condition}) we see that the array is commensurate if and only if there are 2 non-negative integers $n_1$ and $n_2$ with the greatest common divisor (GCD) equal to 1 such that there is a positive L with
\begin{equation}
\begin{array}{ccl}
\frac L2 \sqrt{2x^2 + 2 \sqrt{x^4-4u^4}} &=& 2\pi n_1\\[0.3cm]
\frac L2 \sqrt{2x^2 - 2 \sqrt{x^4-4u^4}} &=& 2\pi n_2.
\end{array}\label{eq:SolutionA4}
\end{equation}
Eq.~\ref{eq:xu} allows for $a_1$ and $a_3$ to be solved in terms of
$x^2$ , $u^2$ , and $a_2^2$.  Without loss of generality, we set $a_2=qs$, where $q$ is coupling strength analogous to Rabi frequency and $s$ an arbitrary real. Then we use Eq.~(\ref{eq:SolutionA4}) to solve for $x^2$ and $u^2$ in terms
of $n_1$, $n_2$, and $s$. The result is
\begin{equation}
\begin{array}{ccl}
a_1 & = & \frac q2 \left(\epsilon_1\sqrt{(n_1+n_2)^2-s^2}+
\epsilon_2\sqrt{(n_1-n_2)^2-s^2}\right)\\[.3cm]
a_2 &=& qs\\[.3cm]
a_3 & = & \frac q2 \left(\epsilon_1\sqrt{(n_1+n_2)^2-s^2}-
\epsilon_2\sqrt{(n_1-n_2)^2-s^2}\right)
\end{array}\label{eq:a1a2a3}
\end{equation}
where $\epsilon_1$ and $\epsilon_2$ in $\{-1,+1\}$.

Note that for any eigenspectrum given by $n_1$ and $n_2$ an infinite number of arrays can be constructed by choosing different $s$ and $\epsilon_{1,2}$.
Not all values of $s$ will give real values for the $a_i$. However, all will give eigenvalues with the ratios $\{\pm n_1, \pm n_2\}$.

An interesting example occurs when we choose degenerate eigenvalues, $n_1=n_2$. Then we are forced
to choose $s=0$ and hence $a_2=0$, meaning that the array decomposes into two 2-waveguide arrays with the same coupling coefficients $a_1=a_3$.

\subsubsection{5-guide array}\label{subsec:A5}
We turn to the commensurability of $A_5(a_1,a_2,a_3,a_4)$ given by
\begin{equation}\label{eq:A5}
A_5(a_1,a_2,a_3, a_4)\equiv
\left[ \begin {array}{ccccc}
0&{\it a_1}&0&0&0\\
\noalign{\medskip}{\it a_1}&0&{\it a_2}&0&0\\
\noalign{\medskip}0&{\it a_2}&0&{\it a_3}&0\\
\noalign{\medskip}0&0&{\it a_3}&0&{\it a_4}\\
\noalign{\medskip}0&0&0&{\it a_4}&0\end {array} \right].
\end{equation}
The central eigenvalue of this array in always $0$. We assume that the other eigenvalues relate to one another as $(\pm n_1$, $\pm n_2$) and repeat the procedure from the previous section. The matrix $A_5$ is commensurate if and only if:
\noindent \underline{If $a_1^2-a_4^2\neq 0$}, there are integers $n_1$, $n_2$ with GCD equal to 1 and real numbers s and t, such that $a_i$ satisfy:
\begin{equation}
\begin{array}{ccl}\label{eq:SolutionA5}
a_1^2 & = & q^2s^2\\[0.0cm]
a_2^2 & = & q^2\left(
\frac{(n_1^2n_2^2-s^2t^2)-(n_1^2+n_2^2-(s^2+t^2))s^2}{t^2-s^2}\right)\\[0.3cm]
a_3^2 & = & q^2\left(
\frac{-(n_1^2n_2^2-s^2t^2)+(n_1^2+n_2^2-(s^2+t^2))t^2}{t^2-s^2}\right)\\[0.3cm]
a_4^2 & = & q^2t^2
\end{array}
\end{equation}
where $s$ and $t$ are arbitrary reals and $q$ is an arbitrary non-zero real.\\
\underline{If $a_1^2-a_4^2= 0$} the solution assumes the form:
\begin{equation}
\begin{array}{ccl}
a_1^2 & = & q^2n_1^2\\[0.3cm]
\left(\begin{array}{c} a_2\\a_3\end{array}\right) &=&
\sqrt{q^2(n_2^2-n_1^2)}\,R_\phi\left(\begin{array}{c} 1\\0\end{array}\right)
\end{array}
\end{equation}
where $q$ is an arbitrary non-zero real and $R_\phi$ is a rotation
by an arbitrary angle $\phi$. In the case of an array mirror-symmetric around the centre ($a_j=a_{n-j}, \, j=1,2$), the eigenvalues are easily calculated to be: $\{0,\pm a_1, \pm\sqrt{2a_2^2+a_1^2}\}$. If $a_1\cdot a_2 \neq 0$, the commensurability condition reduces to:
\begin{equation}
\begin{array}{c}
a_1^2=q^2n_1^2\\[0.2cm]
a_2^2=q^2\frac{n_2^2-n_1^2}{2}.
\end{array}
\label{eq:SolutionM5}
\end{equation}
\underline{If $a_1=0$}, the system becomes a trivially periodic 3-guide array with two outer waveguides remaining uncoupled to the others. If $a_2=0$ the central waveguide is decoupled from two pairs of coupled waveguides at both ends of the array.

For an array with an equidistant energy spectrum, reverse engineering renders solutions from the family of Clebsch-Gordan coupling coefficients. For instance, a symmetric 5-element array with $n_2=2n_1$, can have coupling coefficients $a_1=q n_1$, and $a_2=qn_1\sqrt{\frac{3}{2}}$ that correspond to Zeeman states in $|F=2\rangle$ hyperfine level \cite{PetrovicNJP13}. WGAs with this property have been used to construct optical couplers \cite{PetrovicOL15}.

\subsubsection{Symmetric 7-guide array}\label{subsec:A7}
Now we turn our attention to the real, symmetric, nearest-neighbour interaction with
the mirror symmetry acting on $\mathbb{R}^7$
\begin{equation}\label{eq:A7}
A_7(a_1,a_2,a_3)\equiv  \left[ \begin {array}{ccccccc} 0&\it a_1&0&0&0&0&0
\\ \noalign{\medskip}\it a_1&0&\it a_2&0&0&0&0\\ \noalign{\medskip}0&\it a
_2&0&\it a_3&0&0&0\\ \noalign{\medskip}0&0&\it a_3&0&\it a_3&0&0
\\ \noalign{\medskip}0&0&0&\it a_3&0&\it a_2&0\\ \noalign{\medskip}0&0
&0&0&\it a_2&0&\it a_1\\ \noalign{\medskip}0&0&0&0&0&\it a_1&0
\end {array} \right].
\end{equation}
\noindent The system is commensurate with the eigenvalues that relate to each other as $(0, \pm n_1, \pm n_2, \pm n_3)$ if and only if:\\
\noindent\underline{If a3 ̸= 0:} there are integers $n_1$, $n_2$, and $n_3$ with GCD equal to 1 such that the $a_i$ satisfy:
\begin{equation}
\begin{array}{c}
a_1^2=q^2\left(\frac{n_2^2n_3^2}{n_2^2+n_3^2-n_1^2}\right)\\[0.4cm]
a_2^2=q^2\left(n_1^2-\frac{n_2^2n_3^2}{n_2^2+n_3^2-n_1^2}\right)\\[0.4cm]
a_3^2=q^2\left(\frac{n_2^2+n_3^2-n_1^2}{2}\right)
\end{array}\label{eq:SolutionA7}
\end{equation}
where $q$ is an arbitrary positive real.\\
\noindent\underline{If $a_3=0$}, the array is decomposed into the central waveguide and two 3-waveguide arrays with the period $L=\frac{2\pi}{\sqrt{a_1^2+a_2^2}}$.

\subsubsection{Symmetric 9-element array}\label{subsec:A9}

In the spirit of the previous sections, we solve the inverse problem for a mirror-symmetric 9-guide array with the real, nearest-neighbour interaction acting on $\mathbb{R}^9$
\begin{equation}\label{eq:A9}
A_9(a_1,a_2,a_3,a_4)\equiv
 \left[ \begin {array}{ccccccccc} 0&{\it a_1}&0&0&0&0&0&0&0
\\ \noalign{\medskip}{\it a_1}&0&{\it a_2}&0&0&0&0&0&0
\\ \noalign{\medskip}0&{\it a_2}&0&{\it a_3}&0&0&0&0&0
\\ \noalign{\medskip}0&0&{\it a_3}&0&{\it a_4}&0&0&0&0
\\ \noalign{\medskip}0&0&0&{\it a_4}&0&{\it a_4}&0&0&0
\\ \noalign{\medskip}0&0&0&0&{\it a_4}&0&{\it a_3}&0&0
\\ \noalign{\medskip}0&0&0&0&0&{\it a_3}&0&{\it a_2}&0
\\ \noalign{\medskip}0&0&0&0&0&0&{\it a_2}&0&{\it a_1}
\\ \noalign{\medskip}0&0&0&0&0&0&0&{\it a_1}&0\end {array} \right].
\end{equation}
The system is commensurate if and only if:\\
\noindent\underline{If a3 ̸= 0 and a4 ̸= 0:} there are integers $n_1$, $n_2$, $n_3$, and $n_4$ with GCD equal to 1 such that the $a_i$ satisfy:\\
\begin{equation}
\begin{array}{c}
a_1^2=q^2\left( \frac{n_1^2n_2^2(n_3^2+n_4^2-n_1^2-n_2^2)}
{(n_1^2+n_2^2)(n_3^2+n_4^2-n_1^2-n_2^2)-(n_3^2n_4^2-n_1^2n_2^2)}\right)\\[0.4cm]
a_2^2=q^2\left(  \frac{n_3^2n_4^2-n_1^2n_2^2}{n_3^2+n_4^2-n_1^2-n_2^2}-
\frac{n_1^2n_2^2(n_3^2+n_4^2-n_1^2-n_2^2)}
{(n_1^2+n_2^2)(n_3^2+n_4^2-n_1^2-n_2^2)-(n_3^2n_4^2-n_1^2n_2^2)}\right)\\[0.4cm]
a_3^2=q^2\left(  \frac{(n_1^2+n_2^2)(n_3^2+n_4^2-n_1^2-n_2^2)-
(n_3^2n_4^2-n_1^2n_2^2)}{(n_3^2+n_4^2-n_1^2-n_2^2)}
\right)\\[0.4cm]
a_4^2=q^2\left( \frac{n_3^2+n_4^2-n_1^2-n_2^2}{2}\right).
\end{array}
\end{equation}
\noindent\underline{If $a_4=0$}, the matrix consists of three diagonal blocks two of which correspond to 4-guide arrays and one block is the number 0.\\
\noindent\underline{If $a_3=0$}, there are also three diagonal blocks, that represent 3-guide arrays.

\subsection{WGA design}\label{subsec:WGA_design}

To design realistic commensurate WGAs, we move beyond the discrete approximation and apply transversal mode field analysis. The evanescent-field coupling mechanism renders exponential decay of the coupling coefficients, $a_i$ as defined in (\ref{eq:A4}) and (\ref{eq:A5}), with the interwaveguide separations, $d_{i,i+1}$, given by $a_i=a_0e^{-\alpha d_{i,i+1}}$ where $i$ indexes waveguides. The parameters $a_0$ and $\alpha$ are determined by the refractive index profile of waveguides in the array (assumed all equal), \cite{Yariv}. For commonly used circular and rectangular waveguides they do not depend on interwaveguide separations, which enables the control of WGA eigenvalues by tailoring only $d_{i,i+1}$. For a given separation between any two waveguides, others can be derived using the formula $d_{i+1,i+2}=d_{i,i+1}+\frac{1}{\alpha}\ln\frac{a_{i}}{a_{i+1}}$, \cite{BellecOL12, PetrovicOL15}, without the need to know $a_0$. Since the coupling coefficients are derived from the discrete model, we pay special attention to remain in the domain of its validity. Limitations of the model can be understood by observing transversal mode-field profiles in Fig.~\ref{fig:WGA}. Closely spaced waveguides share one supermode that facilitates beyond-nearest-neighbour coupling. Larger spacing provides better mode confinement to individual waveguides and nearly negligible overlap of modes from the non-nearest waveguides, thus converging towards the discrete model. In all simulations performed here, the waveguide separation-to-radius ratio was close to 5.4. Applicability of the exact formulas for coupling coefficients is confirmed by results in Section \ref{sec:Results}.

All simulated WGAs were composed of identical circular waveguides with the diameter 8.2 $\rm \mu m$, the core refractive index 1.45 and the substrate refractive index 1.445. Input mode into each waveguide was the fundamental eigenmode at 1550 nm. While here it was convenient to choose waveguide parameters typical of telecom single-mode fibres, scaling properties of Maxwell's equations allow for straightforward construction of smaller WGAs, such as those accessible by direct laser writing \cite{SzameitJMPB08} and lithography \cite{lithography}. It is assumed that the initial state is coherent and distributed to the waveguides of the array by an appropriate encoder. The choice of encoder and decoder depends on a particular qubit state used in processor and is not discussed here. The coupling strength $q$, that features in expressions for matrix elements $a_i$ as a free scaling parameter, was set to $1$. Numerical simulations were performed using the finite-difference beam propagation method with transparent boundary conditions \cite{BPM}.

\section{Results}
\label{sec:Results}
By performing numerical simulations based on the exact inverse-problem solutions, we first show that the proposed design procedure guarantees reconstruction of amplitude and phase of all components of the complex input-state vector. Example WGAs are then chosen to illustrate the versatility of solutions and transferrable classical and quantum states. Representations of states are discussed in separate subsections.

\subsection{Phase and amplitude revivals}\label{subsec:Revivals}
\begin{figure}[ht!]
\begin{center}
\begin{tabular}{cc}
\includegraphics[height=4.5cm]{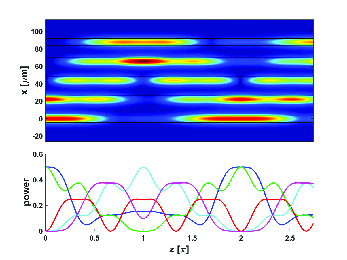}&
\includegraphics[height=4.5cm]{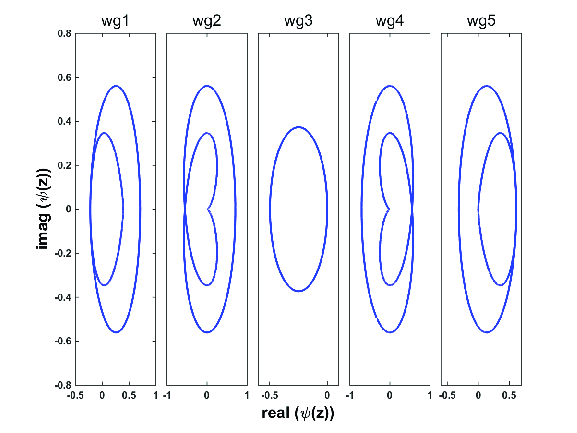}\\
\includegraphics[height=4.5cm]{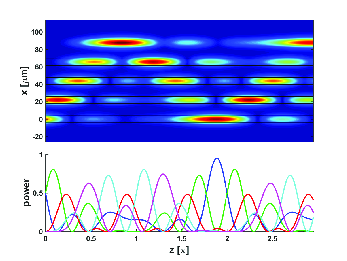}&
\includegraphics[height=4.5cm]{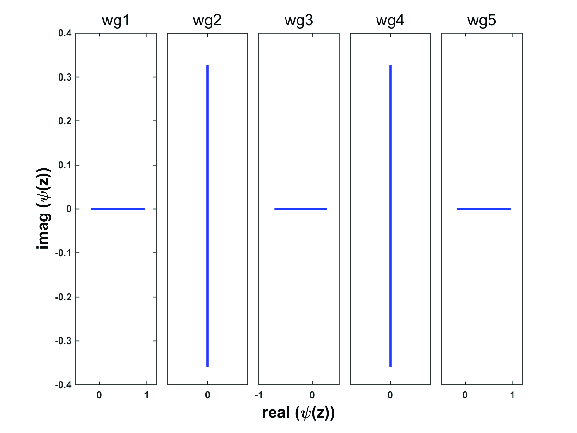}\\
\includegraphics[height=4.5cm]{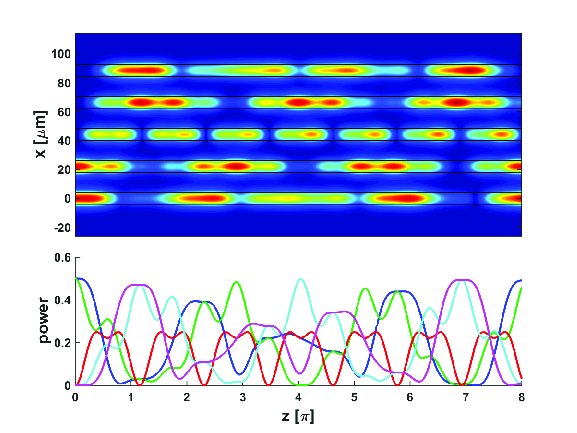}&
\includegraphics[height=4.5cm]{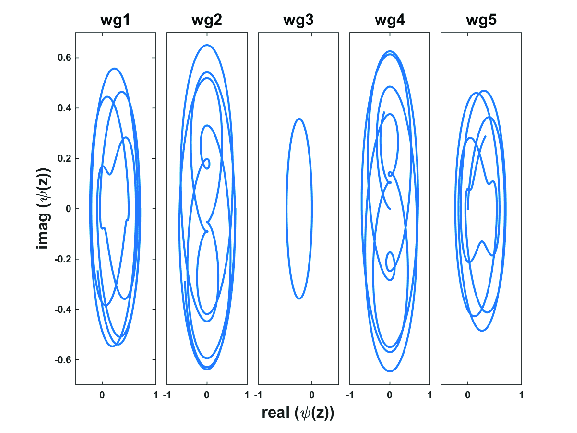}
\end{tabular}
\caption{(a)-(b)Results of numerical simulations of light propagation through a commensurate 5-waveguide WGA with $n_1=2,\, n_2=3$, for different input states: $\psi(0)=\frac{1}{\sqrt{2}}(1,1,0,0,0)$ (a) and $\psi(0)=\frac{1}{\sqrt{2}}(1,e^{-i\pi/2},0,0,0)$ (b). (c) Light propagation of the state $\psi(0)=\frac{1}{\sqrt{2}}(1,1,0,0,0)$ through incommensurate WGA with all coupling coefficients equal and the corresponding eigenvalues $0,\,\pm1,\, \pm\sqrt{3}$. Graphs to the left show intensity profiles (red colour corresponds to the highest intensity) and the power $|\psi(z)|^2$ obtained by solving (\ref{eq:Model}). Colour coding of the exact solutions is blue, green, red, cyan and purple going from the bottom waveguide up. Graphs to the right show paths of $\psi_j(z),\,j=\overline{1,5}$ in the complex plane.}
\label{fig:Coherence}
\end{center}
\end{figure}

To investigate periodicity, we look at the evolution of the state vector in a complex plane. If the evolution is periodic the vector tip renders a finite contour that is swept each period. On the other hand, the state vector with quasi-periodic evolution never repeats the same path but fills in a subspace of the complex plane. The two distinct cases are shown in Fig.~\ref{fig:Coherence}a) and c). Both the amplitude and phase are reproduced at the output. While the periodicity of amplitude is clear from the propagation dynamics, we further confirm the periodicity of phase by showing the full state revivals of states with different relative phases at the input in Fig.~\ref{fig:Coherence}b). Therefore, n-waveguide arrays can be used for the faithful transfer of a vector $\psi(z)$ composed of $n$ complex wavefunctions. The predicted transfer fidelity is $1$. In realistic WGAs fidelity deviates from the ideal, mainly due to the presence of the beyond-nearest-neighbour coupling and the initial mode overlap.

\begin{figure}[ht!]
\begin{center}
\includegraphics[width=8cm]{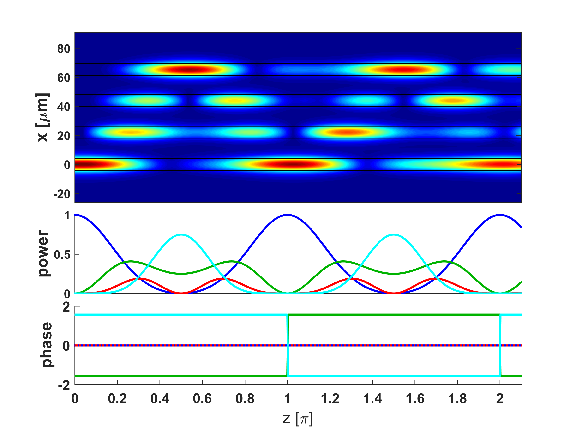}\\
\caption{Results of numerical simulations of light propagation through a commensurate asymmetric 4-waveguide array with eigenvalues set by $n_1=1,\, n_2=3$ and the input vector state $\psi(0)=(1,0,0,0)$. Contour graph shows intensity profiles (red colour corresponds to the highest intensity). Lower graphs show the power $|\psi(z)|^2$ and phase (in radians) obtained by solving (\ref{eq:Model}). Colour coding of the lines is blue, green, red and cyan from the bottom waveguide up.}
\label{fig:Asymmetric}
\end{center}
\end{figure}

The full state revivals can occur in arrays constructed without any additional restrictions on the values of coupling coefficients $a_i$ besides those given in section~\ref{sec:Method}. Hence, a variety of symmetric and asymmetric WGAs can perform the transfer. Example in Fig.~\ref{fig:Asymmetric} shows an asymmetric 4-guide array with commensurate eigenvalues $n_1=\pm 1,\,n_3=\pm 3$, and arbitrarily chosen parameters $s=\sqrt{3}$, $\epsilon_1=1$ and $\epsilon_2=-1$, that render the waveguide separations of $d_{2,3}=0.977 d_{1,2}$ and $d_{3,4}=0.954 d_{1,2}$. Here, amplitude-only revivals occur at each half of the period, while the phase revives at each full period. The fabrication tolerances needed to produce the required interwaveguide separations are discussed in Section~\ref{sec:Discussion}.

\begin{figure}[ht!]
\begin{center}
\begin{tabular}{cc}
\includegraphics[height=5cm]{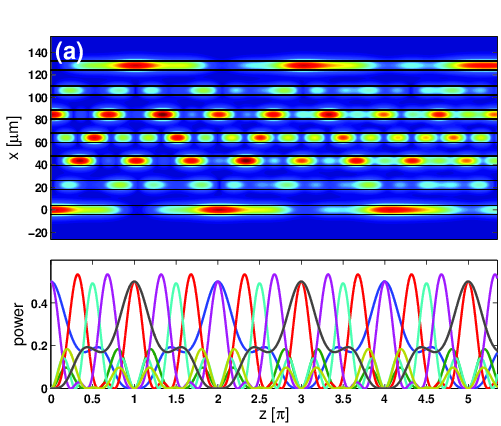}&
\includegraphics[height=5cm]{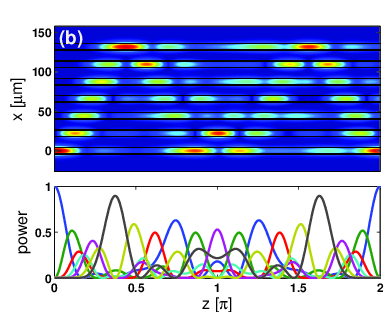}
\end{tabular}
\caption{Results of numerical simulations showing the full coherent transfer through a 7-waveguide array with (a) equidistant eigenvalues $n_1=1,\,n_2=2,\,n_3=3$ for the input vector state $\frac{1}{\sqrt{2}}(1,0,0,0,1,0,0)$, (b) Fibonacci eigenvalues $n_1=5,\, n_2=3,\,n_3=8$ for the input vector state $(1,0,0,0,0,0,0)$. In a) the parallel data transfer is along $z$ and the perfect transfer is along $x$. The full transfer is realized at propagation lengths $m2\pi$ and the \textit{perfect} transfer at lengths $(2m+1)\pi$, where $m=0,1,2,...$. Colour coding of the lines is blue, green, red, cyan, purple, olive green and black going from the bottom waveguide up.}
\label{fig:WGA7}
\end{center}
\end{figure}

We further give two examples of the parallel transfer through 7-waveguide WGAs. The first is a Heisenberg-like chain with equidistant eigenvalues that correspond to the Clebsch-Gordan coupling coefficients of $|F=3\rangle$ hyperfine state, Fig.~\ref{fig:WGA7}a). The second is an array with non-equidistant eigenvalues $n_1=5,\,n_2=3,\,n_3=8$ chosen from the beginning of the Fibonacci sequence.

\begin{figure}[ht!]
\begin{center}
\begin{tabular}{cc}
\includegraphics[height=5cm]{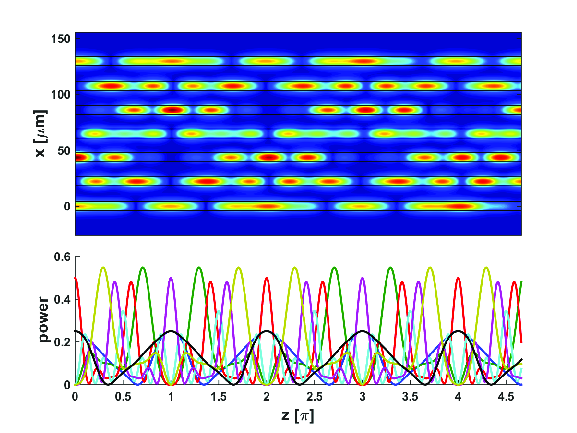}&
\includegraphics[height=5cm]{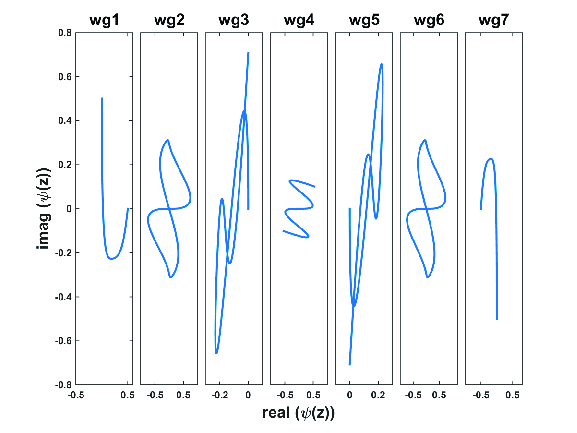}
\end{tabular}
\caption{The transfer of information encoded in three levels of amplitude and phase through a 7-waveguide array with eigenvalues $n_1=1,\,n_2=2,\,n_3=5$. The example input vector state is $\psi(0)=\frac{1}{\sqrt{6}}(1,0,2e^{i\frac{\pi}{2}},0,0,0,e^{-i\frac{\pi}{2}})$. Graphs to the left show intensity profiles (red colour corresponds to the highest intensity) and the power $|\psi(z)|^2$ obtained by solving (\ref{eq:Model}). Colour coding of the exact solutions is blue, green, red, cyan and purple going from the bottom waveguide up. Graphs to the right show paths of $\psi_j(z),\,j=\overline{1,7}$ in the complex plane.}
\label{fig:ternary_WGA7}
\end{center}
\end{figure}

In the absence of noise, each waveguide can transfer a continuum of amplitudes and phases. Figure~\ref{fig:ternary_WGA7} shows the transfer of 3-level phase and amplitude signals through a 7-waveguide array with the eigenvalues $n_1=1,\,n_2=2,\,n_3=5$. Transfer fidelity is comparable to the fidelity for the 2-level transfer from previous examples. In both cases, the non-nearest-neighbour coupling and the mode overlap at the input, together with the noise present in experiment, set the margin that defines the number of levels per waveguide.

From the above examples it is evident that the eigenvalues may, but do not have to, posses any particular regularity apart from the commensurability.

\subsection{Multi-bit transfer in classical optical computer}\label{subsec:Classical}
Amplitude revivals of $n$ wavefunctions enable transfer of $n$ classical bits along the WGA. Information encoding in this case is akin to electronic 0 or 1, where the light intensity below some threshold represents 0 and above some threshold represents 1. Given the short distance of the on-chip transfer, losses and noise are less limiting than the errors due to the non-nearest-neighbour coupling between waveguides. In examples in Fig.~\ref{fig:WGA} and Fig.~\ref{fig:Asymmetric}, initial states can represent classical registers $11000$ and $1000$ that are transferred though a WGA to any integer number of periods.

The revival of relative phases of waveguide modes adds another degree of freedom with respect to the electronic analogue. By defining the phase in one waveguide as a reference and sending the information encoded in the relative phases in other waveguides, additional $n-1$ bits can be transferred. The phase encoding can be performed by any suitable convention that keeps uniqueness of the phase values. As a result, an n-waveguide array can be used to transfer $2n-1$ classical bits encoded in phase and amplitude in parallel.

Finally, in the absence of excessive noise, the commensurate WGAs can support transfer of numbers represented in higher-level logic systems such as ternary, quaternary, etc. A base-m numeral system is accessed by encoding into $m$ intensity or/and $m$ phase levels in each waveguide with the maximum transfer capacity through an n-waveguide array of $m^{2n-1}$. Figure ~\ref{fig:ternary_WGA7} illustrates ternary logic transfer with the parallel amplitude and phase value sampling from the sets ${0,1,2}$ and ${0,\, -\pi/2,\, \pi/}$, respectively. The faithful reproduction of all three levels along a 7-waveguide array enables transfer of a classical register with $3^{13}$ possible words.

\subsection{Qudit and multi-qubit transfer}\label{subsec:Quantum}%

Since a quantum state can be represented as a linear superposition of orthogonal basis vectors, the complex wavefunctions of waveguide modes can be used to transfer complex factors featuring in the superposition, and thus the quantum state. Assuming a basis with vectors $\langle0|_1\langle0|_2\ldots\langle1|_j\ldots\langle0|$ where $j=\overline{1,n}$ indexes waveguides, state $|\psi(z)\rangle$ can be represented by factors $\psi_j(z)=\langle0|_1\langle0|_2\ldots\langle1|_j\ldots\langle0|_n\psi(z)\rangle$. Such an n-dimensional basis accommodates $\log_dn$ qudits with $d$ states. Therefore, an n-waveguide array can be used to transfer $\log_2n$ qubits or a single qudit with $d=n$. WGAs designed in section \ref{sec:Method} can be used to transfer up to 9 complex numbers, hence up to 3 qubits or a single qu$d$it with up to 9 states. For example, the array from Fig.~\ref{fig:Asymmetric} can transfer 2 qubits or a 4-state qudit, and the arrays from Fig.~\ref{fig:WGA7} and Fig.~\ref{fig:ternary_WGA7} can transfer 2 qubits with 3 redundant channels or a 7-state qudit.

In the case of m-qubit transfer by a WGA with the waveguide number $n>2^m$, the apparent redundancy in a number of waveguides allows for the simultaneous transfer of $N-2^m$ ancillary signals. Assuming the tensor-product representation of the Hilbert space suggested in \cite{KiktenkoPLA15}, the same WGA can be used to transfer complete set of states needed for the realization of a universal gate by either qubits or a qudit. Hence, a 5-waveguide array can be used to transfer the input/output of a 5-state qudit gate or an equivalent 2-qubit universal gate with an ancillary bit. By the same token, a 9-waveguide array can transfer a 9-state qudit or 3 qubits and an ancillary bit.

All WGAs proposed here can be used in classical and quantum computers, whereby the nature of computation is determined by the input state. In a quantum computer, only one read-out is possible, at which the state is collapsed. In a classical computer, detection (if non-destructive) can be performed a number of times at different revival lengths along the array.

\subsection{Perfect transfer}\label{subsec:Perfect}

The perfect transfer through linearly coupled arrays of quantum dots \cite{NikolopoulosJP04} and WGAs \cite{ChapmanNatComms16} is characterized by mirroring of a single qubit at the output of the array \cite{KayIJQI10}. The array coupling coefficients are engineered by solving the inverse eigenvalue problem \cite{BosePRL03, BoseCP07, NikolopoulosJP04, PeschelOL98, XiEJPD08, EfremidisOC05} or by direct derivation of Hamiltonians that permit the permutation operation \cite{KostakPRA07}. Reported realizations rely on solutions found in nature in the form of Bloch oscillations \cite{MorandottiPRA99}, atomic spin chains \cite{PetrovicNJP13}, or their emulation by optical lattices \cite{BellecOL12, ChapmanNatComms16}. Designs of new Hamiltonians that enable nearly perfect transfer comprise a modification of the uniform array \cite{KarbachPRA05} and manipulation of coupling coefficients at the ends of the array \cite{WojcikPRA05, LiPRA05, ZhangAP16}.

Here proposed WGAs with equidistant eigenvalues and the output taken at an odd number of half periods act as Heisenberg arrays \cite{ShorePRA79}, supporting the perfect transfer in $x$ direction, see Fig.~\ref{fig:Coherence} and Fig.~\ref{fig:WGA7}a). Moreover, the whole class of symmetric commensurate WGAs whose eigenvalues satisfy parity conditions are capable of state mirroring. Such are 5-waveguide arrays that satisfy (\ref{eq:SolutionM5}) with even $n_2$ ($n_1<n_2$) and 7-waveguide arrays that satisfy (\ref{eq:SolutionA7}) with even $n_1$, odd $n_2$ and odd $n_3$ ($n_2<n_1<n_3$). These solutions may but do not necessarily have equidistant eigenvalues and hence include, but are not limited to, the Heisenberg arrays. Therefore, a subset of symmetric commensurate WGAs proposed here can be used as new exact solutions for the perfect transfer.

For clarity, we stress the difference between the parallel transfer proposed here and the perfect transfer \cite{BellecOL12, ChapmanNatComms16}. The parallel transfer is observed along the waveguide propagation axes ($z$ axis in figures) and the relative phases of input modes are reconstructed at the output. The perfect transfer is observed in the transversal direction (waveguides acting as elements of a linearly coupled array extending along $x$ axis) and the relative phases of output modes are not relevant.

\section{Discussion}\label{sec:Discussion}

\begin{figure}[ht!]
\begin{center}
\begin{tabular}{cc}
\includegraphics[width=7cm]{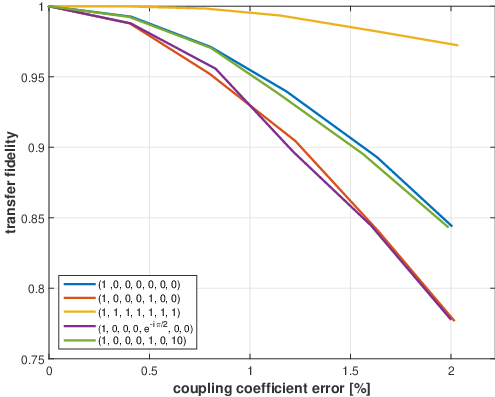}
\end{tabular}
\caption{The transfer fidelity versus standard deviation of normally distributed coupling coefficients. The example WGA has 7 waveguides and $n_1=8,\,n_2=5,\,n_3=13$. The example input states are given in legend.}
\label{fig:Tolerance}
\end{center}
\end{figure}

The interwaveguide separation is a critical parameter in construction of commensurate WGAs. Fabrication errors in interwaveguide separations, $d_{i,i+1}$, induce variation in coupling coefficients, $a_i$, through the exponential dependence given in section~\ref{subsec:WGA_design}. For small deviations, the linear expansion of a coupling coefficient around its design value is a good approximation. To keep the analysis independent of the waveguide parameters, we calculate the sensitivity of the transfer fidelity to small random variations in coupling coefficients, and then relate it to interwaveguide separations and the corresponding tolerance of the available fabrication techniques. The variations in coupling coefficients are modelled by normal distributions centred around their design values with the same standard deviation for all coefficients. To account for the dependence of the transfer fidelity on the state being transferred, we launched different input states, Fig.~\ref{fig:Tolerance}. The transfer fidelity is highly sensitive to deviations of coupling coefficients from their design values. A fidelity over 95\% is guaranteed for deviations as small as 0.8\%. For the WGAs simulated here, such deviation corresponds to the interwaveguide separation tolerance of 25 nm, which is accessible to lithography \cite{lithography}. While the fast progress has been achieved in direct laser writing of waveguides \cite{StreltsovJOSAB02,SzameitJMPB08}, its accuracy is yet to achieve this limit. Nevertheless, diffraction losses as high as 61\% did not prevent observation of the transfer through the laser written WGAs \cite{BellecOL12}, indicating that this technique can be used for proof-of-principle experiments.

In realistic WGAs, coupling to waveguides further than the nearest neighbours is inevitable and introduces low frequency components to the propagation dynamics, thus making the revival length significantly longer (see, e.g., Fig.~4 in \cite{PetrovicOL15}). This effect is observed as slow dephasing over the period characteristic for the nearest-neighbour coupling and does not affect coherence.

The simplicity of the proposed transfer stems from the fact that it relies on the free-evolution of an input state along an array with fixed coupling coefficients, that does not require control of the coupling matrix along the propagation direction. This is achieved at the cost of the discretisation of positions of output ports along the WGA since the state can be faithfully transferred only to an integer number of revival lengths. Another disadvantage is a slow scaling of the link capacity with the number of waveguides. Namely, here adopted model of the faithful transfer assumes decomposition of an input quantum state to $n$ complex coefficients and their straightforward encoding into the amplitude and phase at the input of each waveguide, which introduces unfavourable logarithmic scaling of the number of transferrable qubits with the number of guides. While the above mentioned fabrication techniques allow for the scalability of arrays at low cost, thus partly alleviating the capacity scaling problem, other solutions will be necessary to achieve efficient computation. An adequate solution might be reached by adding a polarization coordinate to the Hilbert space considered here and assuming that the waveguide linear coupling supports polarization degeneracy  \cite{ChapmanNatComms16,CerulloOL02}. This would also open the door to the use of typical polarization-based light qubits.

The equivalency of all waveguides is assumed for convenience, and is not a consequence of any formal requirement. Conveniently, the proposed WGA design procedure is applicable to waveguides of any index profile and geometry that render exponential dependence of coupling coefficients, which holds for all commonly used circular and rectangular waveguides. Particular state encoder and decoder hardware, as well as the corresponding waveguide profile, depend on the physical nature of processor qudits or qubits. Their particular designs are beyond the scope of the current paper.

While the one-dimensional arrays are by construction scalable, it would be interesting to investigate possibilities for their extension to higher dimensions. Although mathematically straightforward, this is not trivial in practise since it is hard to avoid non-nearest-neighbour coupling to the waveguides positioned along diagonals of the waveguide matrix, e.g. coupling of the waveguide ${i,j}$ with waveguides ${\rm i\pm1,\,\rm j\pm1}$. Another extension route may be accessed by coupling the end waveguides to build a circular array \cite{PlenioNJP05, RadosavljevicJOSAB15}.

\section{Conclusions}\label{sec:Conclusions}
We have investigated linearly coupled optical WGAs as means of coherent state transfer on optical chips. The full coherent transfer of amplitude and phase is achieved by reverse engineering of coupling coefficients between the neighbouring waveguides to render the WGA eigenspectrum commensurate. The inverse eigenvalue problem is solved analytically for general arrays with 4 and 5 waveguides and for mirror-symmetric arrays with 7 and 9 waveguides. The analytic solutions provide for the transfer with fidelity 1. These solutions have been applied to construct experimentally accessible commensurate WGAs whose coupling coefficients can be controlled by tailoring interwaveguide separations. Presented WGAs offer a plentitude of possibilities for the parallel transfer of qudits, multiple bits and qubits, and the perfect transfer. Numerical simulations show that the transfer with fidelity above $95\%$ can be supported by scalable WGA architecture realizable by lithographic fabrication techniques. Finding exact solutions for linear arrays with larger number of elements remains a challenge, as do the extensions of commensurate WGAs to higher-dimensional and circular architectures.

\section*{Acknowledgments}
J.J.P. Veerman's research was partially supported by the European Union's Seventh Framework Program (FP7-REGPOT-2012-2013-1) under grant agreement n316165. J. Petrovic's research was supported by the Ministry of Education, Science and Technological Development of Serbia under grant No. III 45010. J. Petrovic acknowledges valuable suggestions from A. Maluckov and F. S. Cataliotti. Authors thank P. Belicev for Fig. 1a.




\section*{References}
\bibliographystyle{elsarticle-num}

\end{document}